\documentclass[pra,twoside,11pt,tightenlines,superscriptaddress]{revtex4}
\usepackage{color}
\usepackage{graphicx,amsbsy,bbm,amsmath,amssymb}
\usepackage{times}
\newcommand{\ie}{{\em i.e.}}
\newcommand{\vp}{{\varrho_p}}
\newcommand{\X}{{\rm X}} 
 \newcommand{\C}{{\rm C}}
\newcommand{\D}{{\rm D}} 
\newcommand{\bfX}{\mathbf X}
\newcommand{\bfY}{\mathbf Y}
\newcommand{\bX}{\overline{\rm X}} \newcommand{\Y}{{\rm Y}}
\newcommand{\bY}{\overline{\rm Y}}
\newcommand{\tru}{{\rm Tr}_1} \newcommand{\trd}{{\rm Tr}_2}
\newcommand{\trud}{{\rm Tr}_{1,2}}
\newcommand{\iid}{\mathbb{I}}
\newcommand{\rmSU}{{\rm SU}}
 \newcommand{\su}{{\sigma_1}}
\newcommand{\sd}{{\sigma_2}} 
\newcommand{\refeq}[1]{Eq.~(\ref{#1})} \newcommand{\ket}[1]{\vert #1 \rangle}
\newcommand{\bra}[1]{\langle #1 \vert} \newcommand{\avg}[1]{\langle#1\rangle}
\begin{document}
\title{Joint measurements on qubits and cloning of observables}
\author{Alessandro Ferraro}
\email{alessandro.ferraro@icfo.es}
\affiliation{ICFO - Institut de Ci\`encies Fot\`oniques, E-08860 Castelldefels 
(Barcelona), Spain} 
\author{Matteo G. A. Paris}
\email{matteo.paris@unimi.it}
\affiliation{Dipartimento di Fisica dell'Universit\`a di Milano, Italia}
\begin{abstract}
Cloning of observables, unlike standard cloning of states, 
aims at copying the information encoded in the statistics of a class of
observables rather then on quantum states themselves. In such a
process the emphasis is on the quantum operation (evolution plus measurement) 
necessary to retrieve the original information. We analyze, 
for qubit systems, the cloning of a class generated by two noncommuting
observables, elucidating the relationship between such a process and
joint measurements. This helps in establishing an optimality criterion
for cloning of observables. We see that, even if the cloning machine
is designed to act on the whole class generated by two noncommuting
observables, the same optimal performances of a joint measurement can
be attained. Finally, the connection with state dependent cloning is
enlightened.
\end{abstract}
\maketitle
\section{Introduction}\label{intro}
It is well known that, in general, the act of measuring a quantum
system necessarily disturb the system itself. A clear exemplification
of this fact is given by joint (or simultaneous) measurements, {\em
i.e.}  measurements in which one tries to simultaneously acquire
information about two noncommuting observables from a single
system. In this case, the possibility to sharply measure (\ie, to
obtain measurement results not affected by more noise than the one
intrinsically present into the system) one observable is forbidden if
a sharp measure of the other one is performed. Nevertheless,
information about both observables can be retrieved, paying the price
of an extra noise in the statistics of the measurement results. An
optimal joint measurement is the one introducing the minimum extra
noise compatible with the laws of quantum mechanics. The first
investigations in this issue can be traced back to the seminal paper
by Arthurs and Kelly \cite{ake}. There, position and momentum
observables have been considered and the minimum extra noise was found
to be four times larger than the intrinsic noise related to the
Heisenberg uncertainty relation. Notice that two points are crucial in
their analysis: i) the equality between the jointly inferred mean
values and the mean values obtained if the measurements were performed
non simultaneously is imposed; ii) the dependence on an {\em a priori}
knowledge of the initial ensemble is needed, in order to adjust the
relative sharpness of the two measurements. The usual implementation
of an Arthurs and Kelly measurement concerns the field of quantum
optics, in which two incompatible quadratures of the field are
considered. Typically, an heterodyne detection scheme (or equivalently
a double homodyne detection) implements the joint measurement, however
also a different strategy can be devised, in the case of fields
described by a coherent state. One can in fact first generate two
(imperfect) copies of the original state, via an optimal cloning
process \cite{cer}, and then measure the two quadratures one for each
clones. Both the strategies perform an optimal joint measurement. This
fact, together with the quite general relationship between cloning and
measurement \cite{cloest}, might lead to think that optimal cloning can
be associated to optimal joint measurements. However, this is not the
case in general. As an example, consider a qubit system and two Pauli
observables, for which again an optimal joint measurement involves an
extra noise four times larger then the intrinsic one
\cite{ago,y82}. An experimental implementation of such a joint
measurement is reported in Ref.\cite{tri}, using an entangled
two-photon state as information carrier from the very beginning.
Then, it is easy to demonstrate that optimal universal cloning for
qubits \cite{buz} does not allows to perform an optimal joint
measurements of the two Pauli observables (see below). This is due to
the fact that cloning machines are in general designed such to
optimize the fidelity between input and output states, not such to
optimize the joint measurement of a specific couple of noncommuting
observables. It is the latter viewpoint that we will instead adopt in
this work. More precisely, we will actually consider a further step,
which can be understood in the spirit of state dependent cloning
\cite{phc}. Recall that the latter cloning strategy, unlike the
universal case, is specifically designed to act only on a particular
class out of the whole Hilbert space of the system ({\em e.g.},
equatorial qubits). Following an analogous strategy, we will design
our process (from now on called {\em cloning of observables}) in order
to clone and possibly jointly measure neither all possible observables
nor two observables only, but rather the class generated by any
linear combination of two observables. More in details, a cloning
machine for the given set of observables is a device in which a signal
qubit interacts with a probe qubit via a given unitary with the aim of
reproducing the statistics of each observable on {\em both} the qubits
at the output, independently of the signal qubit. The optimality of
our cloner will be assessed by investigating the optimality of the
joint measurement (of the two observables generating the class), but
the additional requirement of cloning the whole class of observables
poses a clear distinction between a joint measurement and a cloning of
observables.  Furthermore, unlike a joint measurement, our protocol is
not focused only on the measurement process, which can actually not be
performed. Indeed, our approach is motivated, besides by the
exploitation of a new aspect of the fundamental issue of quantum
cloning, also in view of quantum communication purposes, in a
repeater-like configuration. More specifically, one can address the
transmission of information encoded, rather than in a set of states,
in the statistics of a set of observables, independently of the
quantum state at the input. In the following we will consider set of
observables generated by two Pauli operators, whereas for more
general classes of observables we refer to Ref.~\cite{jpa}.
\par
The paper is organized as follows. In Sec.~\ref{s:CloM} we describe
the general concept of cloning machine for observables, introducing
some basic properties. We point out in Sec.~\ref{s:ncom} the
impossibility of designing a perfect cloning machines for a class of
two noncommuting observables, analogously to the impossibility of
perfectly cloning two nonorthogonal states. Then, an approximate
cloning machine for this case will be introduced. In
Sec.~\ref{s:jointclob} we assess the optimality of such a cloning
machine and compare it to a joint measurement process. We will see
that, even if the cloning machine is designed to act on the whole
class generated by two noncommuting observables, the same
performances of a joint measurement can be attained, when one
restricts the attention to the only two observables generating the
class. Sec.~\ref{s:esco} closes the paper with some remarks.
\section{Cloning of observables}
\label{s:CloM} 
\begin{figure}[h]
\begin{center}
\includegraphics[width=0.7\textwidth]{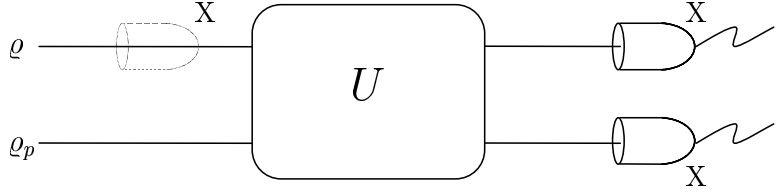}
\caption{Schematic diagram of a cloning machine for observables
$(U,\vp,\bfX)$: a signal qubit prepared in the unknown state
$\varrho$ interacts, via a given unitary $U$, with a probe qubit
prepared in the known  state $\vp$. The class of observables $\bfX$
is cloned if a measurement of any $\X\in\bfX$ on either the two
qubits at the output gives the same statistics as it was measured on
the input signal qubit, {\em independently} on the initial qubit
preparation $\varrho$.} \label{f:machine}
\end{center}
\end{figure}

We consider a device in which a signal qubit (say,
qubit "1") is prepared in the (unknown) state $\varrho$ and then
interacts, via a given unitary $U$, with a probe qubit ("2")
prepared in the known  state $\vp$. For a given class of qubit
observables $\bfX \equiv \left\{ \X(j)\right\}_{j\in{\cal J}}$ where
${\cal J}$ is a subset of the real axis and $\X(j) \in {\cal
L}[{\mathbb C}^2]$, we introduce the concept of cloning as follows
\cite{jpa}. A cloning machine for the class of observables $\bfX$ is
a triple $(U,\vp,\bfX)$ such that
\begin{align}\label{avg}
\bX_1 = \bX \quad \bX_2=\bX \qquad \forall  \varrho \quad \forall\:
\X \in \bfX\:,
\end{align}
where $\bX \equiv
\hbox{Tr}_1 \left[ \varrho\: \X\right]$ is the mean value of the
observable $X$ at the input and
\begin{eqnarray}
 \bX_1 \equiv \hbox{Tr}_{12}\left[ R\: \X \otimes \iid \right]
\quad \bX_2 \equiv \hbox{Tr}_{12}\left[ R\:  \iid \otimes \X\right]
\end{eqnarray}
are the mean values of the same observable on the two output qubits
(see Fig. \ref{f:machine}). The density matrix $R= U\: \varrho
\otimes \vp \:U^\dag$ describes the (generally entangled) state of
the two qubits after the interaction, whereas $\iid$ denotes the
identity operator.
\par
The above definition identifies the cloning of an observable with
the cloning of its mean value. This is justified by the fact that
for any single-qubit observable $\X$ the cloning of the mean value
is equivalent to the cloning of the whole statistics. In fact, any
$\X \in {\cal L}[{\mathbb C}^2]$ has at most two distinct
eigenvalues $\{\lambda_0, \lambda_1\}$, occurring with probability
${p_0,p_1}$. For a degenerate eigenvalue the statement is trivial.
For two distinct eigenvalues we have $\bX = \lambda_1 p_1 +
\lambda_0 p_0$ which, together with the normalization condition
$1=p_0+p_1$, proves the statement.  In other words, we say that the
class of observables $\bfX$ has been cloned if a measurement of any
$\X\in\bfX$ on either the two qubits at the output gives the same
statistics as it was measured on the input signal qubit, {\em
independently} on the initial qubit preparation.
\par
A remark about this choice is in order. In fact, in view of the
duality among states and operators on an Hilbert space, one may
argue that a proper figure of merit to assess a cloning machine for
observables would be a fidelity--like one. This is certainly true
for the $d$-dimensional case, $d>2$, while for qubit systems a
proper assessment can be also made in term of mean--value
duplication, which subsumes all the information carried by the
signal. Furthermore, in view of the possibility to realize a joint
measurement via a cloner of observables, we recall that the Eq.
\refeq{avg} is a standard requirement in such a scenario, as already
recalled in Sec.~\ref{intro}
\par
Before beginning our analysis let us illustrate a basic property of
cloning machines, which follows from the definition, and which will
be used throughout the paper. Given a cloning machine
$\left(U,\vp,\bfX\right)$, then $\left(V,\vp, {\mathbf Y} \right)$
is a cloning machine too, where $V=(W^\dag \otimes W^\dag) \:U
\:(W\otimes \iid)$ and the class ${\mathbf Y}=W^\dag \bfX W$ is formed
by the observables $Y(j)=W^\dag X(j)\,W$, $j\in{\cal J}$. The
transformation $W$ may be a generic unitary. We will refer to this
property to as {\em unitary covariance} of cloning machine. The
proof proceeds as follows. By definition $\bY(j)=\tru[\varrho\:
W^\dag X(j) W] =\tru[W\,\varrho\,W^\dag\,\X(j)]$. Then, since
$\left(U,\vp,\bfX \right)$ is a cloning machine, we have
\begin{align}\label{eq3}
\bY(j) &=\trud[U\,(W\varrho\, W^\dag \otimes \vp)\,U^\dag\,(\X(j)
\otimes \iid)] \nonumber \\ &=\trud[U\,(W\otimes \iid)(\varrho
\otimes \vp)(W^\dag \otimes \iid)\,U^\dag\: (W\otimes W)(\Y(j)
\otimes \iid)(W^\dag\otimes W^\dag)] \nonumber \\
&=\trud[V\,(\varrho \otimes \vp)\,V^\dag\,(\Y(j) \otimes
\iid)]=\bY_1(j)\;.
\end{align}
The same argument holds for $\bY_2(j)$ \cite{note}
\par
Another result which will be used in the following is the
parameterization of a two-qubit transformation, which corresponds to
a $\rmSU (4)$ matrix, obtained by separating its local and
entangling parts. A generic two-qubit gate $\rmSU (4)$ matrix may be
factorized as follows \cite{car}:
\begin{align}\label{SU4}
  U=L_2\,U_{\rm E}\,L_1=L_2\,
\exp\Big[\frac{i}{2}\sum_{j=1}^{3}\theta_j\sigma_j\otimes\sigma_j\Big]
\,L_1
\end{align}
where $\theta_j \in \mathbb{R}$ and the $\sigma_j$'s are the Pauli's
matrices. The local transformations $L_1$ and $L_2$ belongs to the
$\rmSU (2)\otimes \rmSU (2)$ group, whereas $U_{\rm E}$ accounts for
the entangling part of the transformation $U$. In our context,
decomposition (\ref{SU4}), together with unitary covariance of
cloning machines, allows to ignore the local transformations $L_1$,
which corresponds to a different state preparation of signal and
probe qubits at the input. On the other hand, as we will see in the
following, the degree of freedom offered by the local
transformations $L_2$ will be exploited to design suitable cloning
machines for noncommuting observables.
\section{Noncommuting observables}\label{s:ncom}
In order to introduce cloning machines for a class of noncommuting
observables let us consider the specific class $\bfX_{\rm
  nc}=\{x_1\su+x_2\sd\}_{x_1,x_2\in\mathbb R}$. If a cloning machine
$(U,\vp,\bfX_{\rm nc})$ existed, then the mean values as well as the
statistics of any observable belonging to $\bfX_{\rm nc}$ would be
cloned at its output. As a consequence, one would jointly measure any
two non-commuting observables belonging to $\bfX_{\rm nc}$ ({\em
  e.g.}, $\su$ on the output signal and $\sd$ on the output probe)
without any added noise, thus violating the bounds imposed by quantum
mechanics \cite{ake,ago,y82}. Generalizing this argument to any
two--parameter class of noncommuting observables ({\em i.e.}, to any
class $\bfX_{\rm gnc}=\{c\,\C+d\,\D\}_{c,d\in{\mathbb R}}$, with
$\C,\,\D$ generic non-commuting observables), we then conclude that a
cloning machine for a generic two--parameter class of noncommuting
observables does not exist \cite{jpa}. This results resembles the
no--cloning theorem for states, and it can actually be seen as its
counterpart for cloning of observables. Furthermore, it permits a
comparison among cloning machines for observables and for states.  Let
us write the generic input signal as $\varrho = \frac12 (\sigma_0 +
{\mathbf s} \cdot{\boldsymbol \sigma})$, where ${\boldsymbol
  \sigma}=(\sigma_1,\sigma_2,\sigma_3)$, ${\boldsymbol
  s}=(s_1,s_2,s_3)$ is the Bloch vector and $\sigma_0$ is the identity
operator. The state--cloning counterpart of the statement above can
then be obtained by considering the class of observables $\bfX_{\rm
  nc}$: If a cloning machine $(U,\vp,\bfX_{\rm nc})$ existed, then the
components $s_1$ and $s_2$ of Bloch vector $\boldsymbol s$ would be
cloned for any input signal. The same situation occurs in the case of
a two--parameter class generated by any pair of Pauli operators. In
other words, it is not possible to simultaneously copy a pair of
components of the Bloch vector of a generic state, even completely
disregarding the third one \cite{kie,app}. Let us stress the fact
that, even if a perfect cloning of observables could be performed (for
example, in the case of a class generated by commuting observables
\cite{jpa}), correlations between the measurements would exist,
otherwise state estimation limits would be overcame.
\par
Now, a question arises on whether, analogously to state--cloning, we
may introduce the concept of approximate cloning machines, {\em
i.e.} cloning of observables involving added noise. Indeed this can
be done and optimal approximate cloning machines corresponding to
minimum added noise may be found as well. \par An approximate
cloning machine for the class of observables $\bfX$ is defined as
the triple $(U,\vp,\bfX)_{\rm apx}$ such that $\bX_1=\bX/g_1$ and
$\bX_2=\bX/g_2$ {\em i.e.}
\begin{align}
\hbox{Tr}_1\left[ \varrho\: \X\right] = g_1 \hbox{Tr}_{12}\left[ R\:
\X \otimes \iid \right] = g_2 \hbox{Tr}_{12}\left[ R\: \iid \otimes
\X\right] \label{appcl} \:,
\end{align}
for any $\X \in \bfX$. The quantities $g_j$, $j=1,2$ are independent
on the input state and are referred to as the noises added by the
cloning process.
\par
Let us begin by again considering the class $\bfX_{\rm nc}=
\{x_1\su+x_2\sd\}_{x_1,x_2\in\mathbb R}$. By using the decomposition
of a generic $\rmSU (4)$ matrix in \refeq{SU4} one may attempt to
find an approximate cloning machine considering only the action of
the entangling kernel $U_{\rm E}$. Unfortunately, it can be shown
that no $U_{\rm E}$, $g_1$ and $g_2$ exist which realize approximate
cloning for $\vp=\ket{0}\bra{0}$. A further single-qubit
transformation should be introduced after $U_{\rm E}$. In
particular, the unitary $F=i/\sqrt{2}(\sigma_1 + \sigma_2)$ flips
the Pauli matrices $\su$ and $\sd$ ({\em i.e.},
$F^\dag\sigma_{1,2}\,F=\sigma_{2,1}$) and permits the realization of
an approximate cloning machine. Indeed, the unitary $$T=(\iid\otimes
F)U_{\rm nc} \qquad
  U_{\rm nc}=e^{i\frac\theta2 (\su\otimes\su-
    \sd\otimes\sd)}\:,$$
realizes the approximate cloning machine
$(T,\ket{0}\bra{0},\bfX_{\rm nc})_{\rm apx}$ with added noises
\begin{eqnarray}
g_1=\frac1{\cos\theta} \qquad g_2=\frac1{\sin\theta}\:.
\label{addn2}
\end{eqnarray}
In order to prove the cloning properties of $T$ let us start from
the unitary $(\iid\otimes F)U_{\rm E}$, where $U_{\rm E}$ is a
generic entangling unitary of the form given in Eq. (\ref{SU4}).
Then, by imposing approximate cloning for any $\X\in\bfX_{\rm nc}$,
one obtains the following system of Equations:
\begin{subequations}\label{nccm_aux}
\begin{align}
  g_1\trd[(\iid\otimes\vp)\,U_E^\dag\,(\su\otimes\iid)\,U_E]&=\su  \\
  g_1\trd[(\iid\otimes\vp)\,U_E^\dag\,(\sd\otimes\iid)\,U_E]&=\sd  \\
  g_2\trd[(\iid\otimes\vp)\,U_E^\dag\,(\iid\otimes\sd)\,U_E]&=\su  \\
  g_2\trd[(\iid\otimes\vp)\,U_E^\dag\,(\iid\otimes\su)\,U_E]&=\sd
\:.\end{align}
\end{subequations}
System (\ref{nccm_aux}) admits the solution
$\theta_1=-\theta_2=\theta/2$, $\theta_3=0$---{\em i.e.}, $U_E
\equiv U_{\rm nc}$ with $\theta$ free parameter---with
$g_1=1/\cos\theta$ and $g_2=1/\sin\theta$. Notice that other
solutions for the $\theta_{1,2,3}$'s parameters may be found, which
however give the same added noise as the one considered above.
\par
Remarkably, similar cloning machines may be obtained for any class
of observables generated by a pair of operators unitarily equivalent
to $\su$ and $\sd$. In fact, given the two-parameter classes of
noncommuting observables defined as $\bfX_{\rm
V}=\{c\,\C+d\,\D\}_{c,d\in{\mathbb R}}$, with $\C=V^\dag\su V$,
$\D=V^\dag\sd V$ and $V$ generic unitary one has that an approximate
cloning machine is given by the triple
$(U_V,\ket{0}\bra{0},\bfX_{\rm V})_{\rm apx}$, with $U_V=(V^\dag
\otimes V^\dag) (\iid\otimes F)\:U_{\rm nc} \:(V\otimes \iid)$, with
added noises $g_1=1/\cos\theta$ and $g_2=1/\sin\theta$. The
statement easily follows from the fact that
$(T,\ket{0}\bra{0},\bfX_{\rm nc})_{\rm apx}$ is a cloning machine
and from unitary covariance. Similar results hold for any class of
observables unitarily generated by any pair of (noncommuting) Pauli
operators.
\section{Optimality and joint measurements}\label{s:jointclob}
Let us now consider if the approximate cloning machine for
observables introduced above is optimal. In order to assess the
quality and to define optimality of a triple $(U,\vp,\bfX)_{\rm
apx}$ we consider it as a tool to perform a joint measurements of
noncommuting qubit observables \cite{tri}. For example, consider the
cloning machine
 $ (T,\ket{0}\bra{0},\bfX_{\rm nc})_{\rm apx}$
 and suppose to measure $\su$ and $\sd$ on the two qubits at the output.
We emphasize again that the cloning machine $
(T,\ket{0}\bra{0},\bfX_{\rm nc})_{\rm apx}$ clones every observable
belonging to $\bfX_{\rm nc}$, while we are now considering only the
observables $\su$ and $\sd$ which, in a sense, generate the class.
The difference with a joint measurement specifically designed to
measure only a couple of observables is evident. Indeed {\em a
priori} it is not clear at all if a cloning for observables, being
in this sense far more general than a joint measurement, can reach
the minimum disturbance necessary to perform an optimal joint
measurement. Nevertheless, this turns out to be the case, as we will
see in the following. We have that the measured expectation values
of $\su$ and $\sd$ at the output are given by $\avg{\sigma_h}_{\rm
m}=g_h\avg{\sigma_h}$ (with $h=1,2$), where the $\avg{\sigma_h}$'s
are the input mean values. It follows that the measured
uncertainties ($\Delta O=\avg{O^2}-\avg{O}^2$) at the output are
given by $$\Delta_{\rm m}\sigma_h=g_h^2\Delta_{\rm i}\sigma_h$$
where $\Delta_{\rm i}\sigma_h$ denote the intrinsic uncertainties
for the two quantities at the input. Since for any Pauli operators
we have $\sigma_h^2=\iid$ one may rewrites
\begin{align}
\Delta_{\rm m}\su &=\tan^2\theta+\Delta_{\rm i}\su \\
\Delta_{\rm m}\sd &=\cot^2\theta+\Delta_{\rm i}\sd\:.
\end{align}
As a consequence, the measured uncertainty product is given by:
  $$ \Delta_{\rm m}\su\Delta_{\rm m}\sd=\Delta_{\rm i}\su\Delta_{\rm i}\sd
+\cot^2\theta\Delta_{\rm i}\su+\tan^2\theta\Delta_{\rm i}\sd+1\:.$$
Since the arithmetic mean is bounded from below by the geometric
mean we have
  $$\cot^2\theta\Delta_{\rm i}\su+\tan^2\theta\Delta_{\rm i}\sd
  \ge 2\sqrt{\Delta_{\rm i}\su\Delta_{\rm i}\sd}\:,$$
  with the equal sign iff $\Delta_{\rm i}\su=\tan^4\theta\Delta_{\rm
  i}\sd$, then it follows that
  $$\Delta_{\rm m}\su\Delta_{\rm m}\sd\ge
  \left(\sqrt{\Delta_{\rm i}\su\Delta_{\rm i}\sd}+1\right)^2\:.$$
  If the initial signal is a minimum uncertainty state---{\em i.e.},
  $\Delta_{\rm i}\su\Delta_{\rm i}\sd=1$---one finally has that the
  measured uncertainty product is bounded by
  $\Delta_{\rm m}\su\Delta_{\rm m}\sd\ge4$.
  Notice that
  an optimal joint measurement corresponds to have
  $\Delta_{\rm m}\su\Delta_{\rm m}\sd = 4$. In our case this
  is realized when $\theta$ is chosen such that
  $\tan^4\theta=\Delta_{\rm i}\su/\Delta_{\rm i}\sd$. As already
  recalled in Sec.~\ref{intro}, such an {\em a priori} knowledge of
  the initial state is a standard requirement of joint measurements. Therefore, since
 $ (T,\ket{0}\bra{0},\bfX_{\rm nc})_{\rm apx}$
adds the minimum amount of noise in a joint measurement performed on
minimum uncertainty states we conclude that it is an optimal
approximate cloning machine for the class under investigation. An
optimal approximate cloning machine for the more general class
$\bfX_{\rm gnc}$ may be also defined, using the concept of joint
measurement for noncanonical observables \cite{tri}.
\par
Let us now consider the comparison with a joint measurement of $\su$
and $\sd$ performed with the aid of an optimal universal cloning
machine for states \cite{buz}. It is easy to show that the best
result in this case is given by $\Delta_{\rm m}\su\Delta_{\rm
m}\sd=\frac92$, indicating that cloning of observables is more
effective than cloning of states to perform joint measurements (for
the case of three observables see Refs. \cite{hil,pva}). In fact, a
symmetric cloning machine for states shrinks the whole Bloch vector
$\boldsymbol s$ by a factor $\frac23$, whereas a cloning machine for
observables shrinks the components $s_1$ and $s_2$ of $\boldsymbol
s$ only by a factor $1/\sqrt2$ (considering equal noise
$g_1=g_2=\sqrt2$). Notice that such a behavior is different from
what happens in the case of continuous variables, for which the
optimal covariant cloning of coherent states also provides the
optimal joint measurements of two conjugated quadratures \cite{cer}.
This is due to the fact that coherent states are fully characterized
by their complex amplitude, that is  by the expectation values of
two operators only, whereas the state of a qubit requires the
knowledge of the three components of the Bloch vector.
\par
As a final remark we notice that if the requirement of universality
is dropped, then cloning machines for states can be found that
realize optimal approximate cloning of observables. For example, an
approximate cloning machine for the  two--parameter class $\bfX_{\rm
nc}$ can be obtained by considering a phase--covariant cloning for
states \cite{phc}. In order to clarify this point, let us recall
that a phase--covariant cloning machine for states of the form uses
a probe in the $|0\rangle$ state and performs the following
transformation:
\begin{align}
 |0\rangle |0\rangle & \rightarrow  |0\rangle|0\rangle
 \nonumber \\
 |1\rangle |0\rangle & \rightarrow
  \cos\theta|1\rangle|0\rangle+
  \sin\theta|0\rangle|1\rangle)\,,\label{eqc}
 \end{align}
where, in general, $\theta\in[0,2\pi]$. If we now consider the
$\bfX_{\rm nc}$ class it is straightforward to show that Eqs.
(\ref{appcl}) are satisfied for any $\X\in\bfX_{\rm nc}$ using the
machine (\ref{eqc}), with the optimal added noises given by Eq.
(\ref{addn2}). This can be intuitively understood by considering
that a phase covariant cloning machine extracts the optimal
information about states lying on the equatorial plane of the Bloch
sphere, which in turn include the eigenstates of $\sigma_1$ and
$\sigma_2$. Nevertheless, notice that there exist optimal approximate 
cloning machines for observables which do not coincide with
phase--covariant cloning for states. As an example, considering a
cloning machine with added noise $g_1=-1/\cos\theta$ and
$g_2=-1/\sin\theta$ one would still obtain optimal cloning of
observables, even if such a machine could not perform a phase
covariant cloning of states. This remark shows that, in general, one
cannot simply transfer the results for state cloning into the cloning
of observables case.
\section{Conclusions}\label{s:esco}
We have analyzed in details the cloning for classes of observables 
with focus on classes generated by two noncommuting
observables.  We have elucidated the relationship between cloning of observables 
and joint measurements and shown that  even if the cloning machine
is designed to act on the whole class generated by two noncommuting
observables, the same optimal performances of a joint measurement can
be attained. 
\section*{Acknowledgments}
This work has been supported by MIUR through the project
PRIN-2005024254-002. AF acknowledges the financial support from
Universit\`a di Milano under Grant ``Borse di perfezionamento
all'estero''.



\begin{thebibliography}{99}
\bibitem{note} Notice that a cloning 
machine of the form  $(V,\vp,\bfY)$ with $V = (W^\dag \otimes A^\dag) U (W
\otimes \iid )$, where $A$ is an arbitrary unitary matrix also assures that
$\bY (j)=\bY_1(j)$. On the other hand, in order to 
have $\bY (j)=\bY_2(j)$ at the same time, the relation $A=W$
should hold.
\end{thebibliography}
\end{document}